# Robust negative longitudinal magnetoresistance and spin-orbit torque in sputtered Pt$_3$Sn topological semimetal

Delin Zhang,[1*†] Wei Jiang,[1*] Hwanhui Yun,[2] Onri Jay Benally,[1] Thomas Peterson,[3] Zach Cresswell,[1] Yihong Fan,[1] Yang Lv,[1] Guichuan Yu,[4] Javier Garcia Barriocanal,[4] Przemyslaw Swatek[1], K. Andre Mkhoyan,[2] Tony Low,[1†] Jian-Ping Wang[1,2,3†]

[1]Department of Electrical and Computer Engineering; [2]Chemical Engineering and Materials Science; [3]School of Physics and Astronomy; [4]Characterization Facility, University of Minnesota, Minneapolis, MN 55455, USA.

*These authors contributed equally to this work.
†Corresponding authors: jpwang@umn.edu (J.P.W.), tlow@umn.edu (T.L.) and dlzhang@umn.edu (D.L.Z.).

**Contrary to topological insulators, topological semimetals possess a nontrivial chiral anomaly that leads to negative magnetoresistance and are hosts to both conductive bulk states and topological surface states with intriguing transport properties for spintronics. Here, we fabricate highly-ordered metallic Pt$_3$Sn and Pt$_3$Sn$_x$Fe$_{1-x}$ thin films via sputtering technology. Systematic angular dependence (both in-plane and out-of-plane) study of magnetoresistance presents surprisingly robust quadratic and linear negative longitudinal magnetoresistance features for Pt$_3$Sn and Pt$_3$Sn$_x$Fe$_{1-x}$, respectively. We attribute the anomalous negative longitudinal magnetoresistance to the type-II Dirac semimetal phase (pristine Pt$_3$Sn) and/or the formation of tunable Weyl semimetal phases through symmetry breaking processes, such as magnetic-atom doping, as confirmed by first-principles calculations. Furthermore, Pt$_3$Sn and Pt$_3$Sn$_x$Fe$_{1-x}$ show the promising performance for facilitating the development of advanced spin-orbit torque devices. These results extend our understanding of chiral anomaly of topological semimetals and can pave the way for exploring novel topological materials for spintronic devices.**

Topological materials provide a promising platform for exploring intriguing physics and designing new materials (1-3). Given the unique chiral topological surface/edge states in topological materials (4-6), they have been proposed as novel spin-orbit torque (SOT) memory devices with large spin-torque efficiency ($\theta_{SH}$) (7-12). Most notable technological advancements are the demonstration of $\theta_{SH}$ larger than 10 with the switching current density ($J_c$) lower than $10^6$ A/cm$^2$ in sputtered topological insulators at room temperature (10-12). Most recent efforts, however, have shifted to the exploration of novel topological semimetals, which possess the exotic physics of the topological bulk states with conduction and valence bands touching at points (WSM/DSM) or lines (nodal line semimetals) (6,13-16), which could benefit more energy-efficient and industry-compatible SOT devices. Recently, a relatively large $\theta_{SH}$ and SOT magnetization switching through several topological semimetals have indeed been experimentally demonstrated (17-24), hence ushering in their exploration for topological spintronic applications.

One of the most intriguing transport phenomena in topological semimetals is the chiral (Adler-Bell-Jackiw) anomaly, which predicts the transfer of Weyl Fermions with opposite chirality in the presence of parallel electric and magnetic fields (13~16,25~28). Negative longitudinal magnetoresistance (NLMR) is one of the manifestations of chiral anomaly and commonly considered as an experimental signature of topological semimetals (29-34), as reported in Na$_3$Bi (13,27), TaAs (14-16), WTe$_2$ (35), ZrTe$_5$ (36), Cd$_3$As$_2$ (37,38), TaP (39), among many others. However, the exact origin of NLMR is still an open debate as there are alternative thin film material systems showing NLMR absence of chiral anomaly, such as topological insulators or disordered semiconductors, whose origin are attributed to Berry curvature induced anomalous velocity and its derivative orbital moment or Zeeman effect on percolating current pathways in disordered bulk (40-46).

Recent theoretical works also predict a linear NLMR in time-reversal symmetry (TRS) breaking WSM instead of quadratic NLMR in the TRS counterpart based on Onsager's relations, where electric current depends linearly on the magnetization (31-33). Contrary to typical magneto-transport experiments, the NLMR associated with chiral anomaly in a thin film semimetal relies only on the in-plane magnetic field. This effect should be accompanied by additional contributions of the conventional positive magnetoresistance (MR) due to carrier localization induced by the out-of-plane magnetic field. Systematic angle dependent (both in-plane azimuthal and out-of-plane polar angles) magneto-transport study of topological semimetals with/without TRS breaking would allow us to validate and reconcile these different contributions.

Recently, binary Pt-Sn alloys have been identified as a new family of topological materials, which have five known stable phases of different compositions that show rich topological properties. Besides the $PtSn_4$ single crystal, which has already been demonstrated experimentally to be a nodal line Dirac semimetal (47), a more interesting $Pt_3Sn$ alloy has been predicted theoretically to be a promising three-dimensional weak topological insulator hosting type-II Dirac fermion (48). However, there is no experimental investigation of its topological properties and application in spintronic devices to-date. Meanwhile, fabrication process of a high-ordered and industrial-compatible $Pt_3Sn$ is less challenging compared to some topological materials (e.g. $Bi_2Se_3$ (10) and $WTe_2$ (20,23)).

In this work, we successfully fabricate highly ordered Pt$_3$Sn with/without seed layers and Fe-doped Pt$_3$Sn (Pt$_3$Sn$_x$Fe$_{1-x}$) thin films through sputtering deposition that provides seamless integration with the industry development of memory devices for CMOS technology integration. The Pt$_3$Sn with/without seed layers and Pt$_3$Sn$_x$Fe$_{1-x}$ samples show a surprising robust quadratic and linear NLMR, respectively. These results are consistent with the topological semimetal phases of the Pt$_3$Sn and Pt$_3$Sn$_x$Fe$_{1-x}$ samples with and without TRS, respectively, which is further corroborated by our DFT calculations. Meanwhile, both azimuthal and polar angular sweepings of the magnetic field reveal a NLMR behavior that is commensurate with the phenomenon of chiral anomaly and can be reliably reproduced within a simple model. Furthermore, we calculate the spin Hall conductivity to be ~ $4.34 \times 10^5$ $\hbar/2e$ $(\Omega \cdot m)^{-1}$ for Pt$_3$Sn, which is two times larger than that of DSM PtTe$_2$ ($0.2$-$2 \times 10^5$ $\hbar/2e$ $(\Omega \cdot m)^{-1}$) and WSM WTe$_2$ ($2.04 \times 10^5$ $\hbar/2e$ $(\Omega \cdot m)^{-1}$) (21,49). $\theta_{SH}$ of the Pt$_3$Sn and Pt$_3$Sn$_x$Fe$_{1-x}$ thin films is evaluated to be ~ 0.4 and 0.38, respectively, as characterized by spin-torque ferromagnetic resonance (ST-FMR) measurement, which is several times larger than that of PtTe$_2$ (~ 0.1) (21) and WTe$_2$ (~ 0.2) (19) with the same thickness.

**Crystalline structure of Pt$_3$Sn and Pt$_3$Sn$_x$Fe$_{1-x}$**

The crystallinity and microstructure are investigated for the Pt$_3$Sn, Pt$_3$Sn$_x$Fe$_{1-x}$, Pt-seeded and Mo-seeded Pt$_3$Sn thin films deposited on (001) single-crystal MgO substrates with substrate heating at 350 °C. Figure 1(a) shows the crystalline structure of Pt$_3$Sn and Pt$_3$Sn$_x$Fe$_{1-x}$ samples by x-ray diffraction (XRD). We can clearly observe the (111) textured growth for Pt$_3$Sn and Pt$_3$Sn$_x$Fe$_{1-x}$ samples with the (111) peak, which is like the Pt reference sample on MgO substrate. We also observe a small (002) peak for both samples, indicating that these $Pt_3Sn$ thin films have (002) orientated grains only in some regions while the dominant texture is along the (111) direction. To further confirm the crystalline orientation, we carried out reciprocal space mapping (RSM) measurements, in units of the MgO lattice (4.212Å), by XRD, as shown in Fig. 1(b). Both (111) and (002) diffraction patterns can be seen for $Pt_3Sn$ and $Pt_3Sn_xFe_{1-x}$ samples, confirming high-ordered phase. For $Pt_3Sn$ thin films with seed layers, Pt seed layer can maintain the (111) texture with certain (002) orientated grains, the same as $Pt_3Sn$ and $Pt_3Sn_xFe_{1-x}$. However, a Mo seed layer induces the (002) texture, as shown in Figs. S1(a) and S1(b) in Supplemental Information (SI). To further investigate the microstructure and chemical composition of the samples, scanning transmission electron microscopy (STEM) measurements were conducted for $Pt_3Sn$, $Pt_3Sn_xFe_{1-x}$, and Mo-seeded $Pt_3Sn$ samples (Figs. 1(c), 1(d), and Fig. S2). From the atomic-resolution STEM images, it was confirmed that both $Pt_3Sn$ and $Pt_3Sn_xFe_{1-x}$ samples show primarily (111) textured grains with in-plane twist between them as well as some grains with a (002) texture (Fig. 1(c)). Mo-seeded $Pt_3Sn$ sample exhibits only (002) growth (see Fig. S2 for details). STEM-energy dispersive X-ray (EDX) analysis was performed (Fig. 1(d)). Atomic-resolution EDX elemental maps show that the atomic positions of Sn and Fe are overlapping, which directly demonstrates that the Fe atoms are located at the Sn sites. The Fe atoms substituting Sn atoms is also evidenced from comparison of two EDX spectra from $Pt_3Sn_xFe_{1-x}$ and $Pt_3Sn$ samples (see Fig. S2(b)), where a relative increase of Fe $K$ peaks and decrease of Sn $L$ peaks in $Pt_3Sn_xFe_{1-x}$ can be seen.

**Robust negative magnetoresistance of $Pt_3Sn$ and $Pt_3Sn_xFe_{1-x}$**

To investigate the topological properties, the $Pt_3Sn$ and $Pt_3Sn_xFe_{1-x}$ samples were patterned into Hall bar devices with 12-μm width and 144-μm length by using an optical lithography process. The electric-transport properties were tested by a physical property measurement system, as illustrated in Fig. 2(a). The resistivity ($\rho_{xx}$) is measured and calculated to be 162 μΩ.cm and 114 μΩ.cm for $Pt_3Sn$ and $Pt_3Sn_xFe_{1-x}$ at room temperature, respectively (see Fig. S3(a) in SI). With decreasing temperature, $\rho_{xx}$ exhibits a metallic behavior, reaching at 1.9 K a residual value $\rho_0$ of about 83 μΩ.cm and 79 μΩ.cm for $Pt_3Sn$ and $Pt_3Sn_xFe_{1-x}$, respectively, due to carrier scattering with impurities or lattice defects. The residual resistivity ratio RRR = $\rho(300\ K)/\rho(0)$ ~1-2 signals high quality of the studied thin film materials. Meanwhile, the magnetoresistance ($MR_{xx}$) vs. external magnetic field ($H_{ext}$) and Hall resistivity ($R_{xy}$) vs. $H_{ext}$ of these $Pt_3Sn$ and $Pt_3Sn_xFe_{1-x}$ Hall bar devices were measured at 1.9 K for different angles $\theta$ and $\varphi$ [$\theta$ represents the out-of-plane polar angle between $H_{ext}$ and the z-axis; $\varphi$ denotes the in-plane azimuthal angle between $H_{ext}$ and the x-axis (see Fig. 2(a)) where the electric current ($I_c$) is applied along the x-axis], as plotted in Figs. 2(c)-2(f). As shown in Fig. 2(c), we can clearly see that the measured $MR_{xx}$ is surprisingly negative when $\theta < 15^o$ and becomes positive while $\theta > 15^o$ for the $Pt_3Sn$ Hall bar device with a wide range of $H_{ext}$.

This behavior can be ascribed to the presence of at least two competing contributions, whose strength is associated with the polar angle $\theta$ between $I_c$ and $H_{ext}$. Positive contribution of MR is commonly observed in metallic systems that depend only on out-of-plane $H_{ext}$, which can be easily understood from the electron localization induced by magnetic cyclotron orbits (50). On the other hand, negative contribution of MR in 2-dimentional electron gas (2DEG) metallic system is rare, albeit only observed in semiconductor system (41,42). We note that there exist various explanations for the NMR effect, such as the suppression of spin fluctuation under $H_{ext}$, or the weak localization effect, however, most of those effects cannot explain the experimentally observed angle dependence of the NMR (more discussion in Supplementary Note 3). A defining characteristic of the chiral anomaly in topological semimetals is that it only relies on the magnetic field component parallel to the applied electric field ($H_{ext} \parallel I_c$) (29,51). To distinguish the contribution between these two components, we further measured $MR_{xx}$ of the $Pt_3Sn$ Hall bar devices with in-plane rotating $H_{ext}$ with different $\varphi$, as plotted in Fig. 2(d). We can clearly observe NLMR for all azimuthal angles $\varphi$ whose magnitude decreases when $\varphi$ changes from 0 degree to 90 degree. Such robust NLMR phenomenon unequivocally confirms the anomaly contribution from the topological semimetal.

To further investigate the intriguing phenomena of $Pt_3Sn$ system, we tested the transport properties of Pt-seeded $Pt_3Sn$ and Mo-seeded $Pt_3Sn$ that possess different orientation with high crystallinity, as shown in Figs. S3(b)-S3(e). Pt-seeded $Pt_3Sn$ and Mo-seeded $Pt_3Sn$ have similar nontrivial NLMR behavior as that of $Pt_3Sn$ without a seed layer. There is only a slight shape difference in the $MR_{xx}$ curve among $Pt_3Sn$ samples, which can be attributed to sample variations such as doping or crystallinity. It is rather enthralling to observe such robust NLMR in our $Pt_3Sn$ thin films regardless of their different orientations with/without seed layers, strongly suggesting the existence of robust topological semimetal states.

Meanwhile, we also investigated the topological properties of $Pt_3Sn$ doped by magnetic element ($Pt_3Sn_xFe_{1-x}$). With a small amount of Fe dopant (3.8%), $Pt_3Sn_xFe_{1-x}$ presents weak ferromagnetic properties below 25 K (see Supplementary Note 2 and Fig.

S4 in SI). Surprisingly, the MR measurement of $Pt_3Sn_xFe_{1-x}$ with rotating $H_{ext}$ (both $\theta$ and $\varphi$) reveals a very different behavior compared to $Pt_3Sn$. As summarized in Figs. 2(e) and 2(f), we observe a nearly angle-independent negative MR phenomenon with a clear linear $H_{ext}$ dependence. The negative MR shows the linear behavior within the applied $H_{ext}$, which could potentially become non-linear when the magnetic field is larger (52). Unlike positive linear MR that is more commonly observed, linear NLMR has been proposed for WSMs with broken TRS (31-33), see more discussion in SI. In addition, the results of zoomed-in $MR_{xx}$ vs. $H_{ext}$ curves and Hall resistance ($R_{xy}$) vs. $H_{ext}$ curves measured at 1.9 K are presented in Fig. S5, in which the weak anti-localization behavior is clearly observed for these $Pt_3Sn$ and $Pt_3Sn_xFe_{1-x}$ samples. Furthermore, the $Pt_3Sn$ samples show the typical Hall Effect, except for $Pt_3Sn_xFe_{1-x}$ with the anomalous contribution, which can be associated to anomalous Hall effect (AHE) (see Figs. S5(e)-S5(i)). The AHE of $Pt_3Sn_xFe_{1-x}$ can be related to a spin-split band structure due to magnetic dopants, and thus further supports the presence of topologically nontrivial electronic structure in our materials.

**Physical origin of robust NLMR**

To better understand the robust NLMR phenomena observed in $Pt_3Sn$ and $Pt_3Sn_xFe_{1-x}$, a three-resistor model is applied to fit the measured MR results as shown in Fig. 2(b). The resistance of trivial metallic state can be well described by the Drude model $R_c(H_{ext}) = R_{c,0}[1 + \alpha(H_{ext} \cdot sin(\theta))^2]$, contributing to positive MR when $I_c \perp H_{ext}$. While that of the topological semimetal states is assumed to be $R_{SM}(H_{ext}) = R_{SM,0}[1 + \beta(H_{ext} \cdot cos(\theta)cos(\varphi))^2]$, inducing negative MR for $I_c \parallel H_{ext}$ (see details in Supplementary Note 3 and Fig. S6). $R_{c,0}$ and $R_{SM,0}$ represent the initial resistance without $H_{ext}$, which can be extracted from experimental measurements. The total resistance can be simplified using a three-resistor model with one conventional resistor each connected in series and in parallel with $R_{SM}$ that is described as

$$R(H_{ext}) = 1/[\frac{1}{(R_{SM}+R_c^s)} + \frac{1}{R_c^p}]$$ (45). We can easily see that with the increase of $R_{SM}$, the

system has higher tendency to yield negative MR [see Fig. S7(a)], while with the increase of either $R_c^s$ or $R_c^p$, the total MR tends to be more positive [see Figs. S7(a) and S7(c)].

This model captures the experimental $\varphi$-dependent MR feature of Pt$_3$Sn. For the in-plane $\varphi$-dependent measurement, there will be only contributions from $R_{SM}$, whose sign remains negative but magnitude changes with the angle between $I_c$ and $H_{ext}$, as shown in the insert of Figs. 2(c)-2(d) and S8. However, unlike the ideal theoretical model, the experimental MR does not disappear even for $I_c \perp H_{ext}$, possibly due to the polycrystalline nature of the sputtered Pt$_3$Sn that hosts various pairs of Weyl fermions along different directions. Additionally, certain weak localization and weak anti-localization effects, that may cause the deviation between theory and experiments, are not considered. Contrary to Pt$_3$Sn, Pt$_3$Sn$_x$Fe$_{1-x}$ exhibits almost angle-independent linear negative MR behavior [shown in Figs. 2(e) and 2(f)]. Such distinction is possibly related to the distinct nature of the Weyl nodes of TRS-broken WSMs. In time reversal topological semimetal systems, the Weyl pairs are oriented along certain spatial orientations, while the Weyl pairs are locked to $H_{ext}$ in TRS-broken WSMs (see Supplementary Note 4 in SI).

Meanwhile, Pt$_3$Sn$_x$Fe$_{1-x}$ sample shows relatively low crystal quality that have different crystalline orientations, as suggested from Figs. 1(a) and S2(b). Hence, we attribute the azimuthal angle independence to the random crystalline orientations. On the other hand, linear NLMR has indeed been predicted for TRS-broken WSMs when the type-I Weyl nodes are further tilted to form a one-dimensional chiral anomaly (31-33), which agrees with our DFT calculations (see Supplementary Note 4 in SI). Therefore, we fitted the experimental results using a linear model $MR \cong \alpha + \beta H_{ext}$, which perfectly reproduces experimental results, as shown in the insert of Figs. 2(e) and 2(f) (see details in Supplementary Note 3).

To explore the physical origin of the NLMR behavior, we carried out first-principles calculations of $Pt_3Sn$ and $Pt_3Sn_xFe_{1-x}$ (see Figs. 3(a) and 3(d) for corresponding crystalline structures). The band structure of pristine $Pt_3Sn$ is shown in Fig. 3(b). For the pristine phase without SOC, Dirac nodes can be clearly seen at Γ and R points, which become completely gapped when considering SOC, suggesting the topological insulator phase, as also confirmed from our topological edge state calculations (see Fig. S9 in SI). Note that though $Pt_3Sn$ has features of topological insulators, there is a type-II Dirac node and significant bulk states appearing near the Fermi level around R point (Fig. 3(c)), leading to the formation of a "weak topological insulator" (WTI) or topological semimetal (48). The coexistence of topological surface states, Dirac fermion, and metallic bulk states in $Pt_3Sn$ constitutes a physical picture as it is consistent with our previous analysis of the competing contributions to MR that are associated with the angle between electric current and magnetic field. The Dirac fermions contribute to the NLMR due to chiral anomaly while the metallic bulk states contribute to positive MR. It is important to mention that although the Dirac node is not located exactly at the Fermi level, theories have demonstrated that NLMR is robust despite departure from the ideal semimetal systems (34).

The calculated band structure of $Pt_3Sn_xFe_{1-x}$ in the TRS-broken scenario is shown in Fig. 3(e). The band structure without SOC shows a clear spin-splitting due to magnetic doping with each spin channel (spin up (black curve) and spin down (red curve)) hosting a set of bands identical to that of pristine $Pt_3Sn$. When SOC effect is considered, one can clearly see the formation of various pairs of Weyl nodes near the R (see Fig. 3(f)) and Γ points, confirming the formation of TRS-broken WSM phase. Therefore, we have established the topological transition between topological Dirac semimetal and WSM through time-reversal symmetry breaking for the $Pt_3Sn$ and demonstrate the robustness of topological semimetal states against magnetic doping. Considering the possibility of small perturbations of strain or structural defects due to lattice mismatch during sputtering using various seed layers, we also calculated $Pt_3Sn$ with different structural variations, which shows the robustness of topological semimetal states against structural perturbation (see Supplementary Note 4 and Fig. S10 in SI).

**Spin torque efficiency ($\theta_{SH}$)**

After confirming the topological features, we investigate $\theta_{SH}$ of the $Pt_3Sn$ and $Pt_3Sn_xFe_{1-x}$ samples utilizing the ST-FMR technique (53,54). The schematic of the sample stack and testing configuration are illustrated in Figs. 4(a) and 4(b). To precisely evaluate the spin torque efficiency, we fix the thickness of the $Pt_3Sn$ layer and change the thickness of the CoFeB layer (3.0 ~ 6.0 nm). Figures 4(c) and 4(d) show the room-temperature ST-FMR spectra of $Pt_3Sn$ (10.0 nm)/CoFeB (5.0 nm) and $Pt_3Sn_xFe_{1-x}$ (10.0 nm)/CoFeB (5.0 nm) devices, respectively, excited at microwave frequency of 9 GHz. The experimental data (black) is fitted to separate the contribution of symmetric Lorentzian (blue) and antisymmetric Lorentzian (pink) curves. The ($\tau_{FL}+\tau_{Oe}$)/$\tau_{DL}$ vs. $t_{CoFeB}$ is plotted in Figs. 4(e) and 4(f) for $Pt_3Sn$ and $Pt_3Sn_xFe_{1-x}$, respectively, where the slope contains the information for damping-like torque ($\tau_{DL}$), the intercept contains the information for field-like torque ($\tau_{FL}$) and the Oersted field contribution ($\tau_{Oe}$) (see Supplementary Note 5).

The thickness-dependent measurement estimates the $\theta_{SH}$ more reliably by determining the slope of the ($\tau_{FL}+\tau_{Oe}$)/$\tau_{DL}$ ratio over film thickness in Figs. 4(e) and 4(f), as shown by equation $\frac{J_s}{J_c} = \frac{e\mu_0 M_s}{\hbar} \times \frac{\tau_{AD}}{\tau_{Oe}/(t_{CoFeB}d_{Pt_3Sn})}$ (53). From linear curve fitting of Figs. 4(e) and 4(f), the $\theta_{SH}$ of $Pt_3Sn$ and $Pt_3Sn_xFe_{1-x}$ is estimated to be 0.4 and 0.38, respectively. These values are larger than that of the Pt reference ($\theta_{SH} \sim 0.1$) with the same experimental process and testing method (See Fig. S11). Such high $\theta_{SH}$ could originate from the high spin Hall conductivity of the $Pt_3Sn$. A spin Hall conductivity up to $\sim 4.34 \times 10^5$ $\hbar/2e$ $(\Omega \cdot m)^{-1}$ for $Pt_3Sn$ is predicted, which is two times larger than that of WSM $WTe_2$ (See Fig. S12). We note that there is no significant difference of the SOT efficiency between $Pt_3Sn$ and $Pt_3Sn_xFe_{1-x}$, which could possibly be due to the small change of the band structure between the two (more discussion in Supplementary Note 5).

We conclude that the high crystallinity, industrial-compatible topological semimetals, $Pt_3Sn$ and $Pt_3Sn_xFe_{1-x}$, have been fabricated via the sputtering method for the advanced SOT devices. In combination with first-principles calculation and three-transistor model study, we thoroughly study the robust quadratic and linear NLMR features of $Pt_3Sn$ and $Pt_3Sn_xFe_{1-x}$ thin films, which can not only extend the understanding of chiral anomaly in sputtered topological semimetal systems, but also demonstrate the feasibility to design and/or control the topological properties through seed layers or dopants. Meanwhile, the promising SOT performance of the $Pt_3Sn$ and $Pt_3Sn_xFe_{1-x}$ thin films can inspire us to explore more novel topological semimetals for practicable spintronic applications.

## MATERIALS AND METHODS

### Sample preparation and characterization

The $Pt_3Sn$ and $Pt_3Sn_xFe_{1-x}$ samples studied in this work were prepared on single crystal (001) MgO substrates by magnetron sputtering under an ultrahigh vacuum (base pressure < $5.0 \times 10^{-8}$ Torr). The $Pt_3Sn$ thin films were deposited using $PtSn_4$ and Pt targets, and the $Pt_3Sn_xFe_{1-x}$ thin films are prepared using $PtSn_4$, Pt and Fe targets with substrate temperature of 350 $^o$C. The Pt thin film was also grown with the same experimental condition as a reference. The MgO (6.0 nm)/Ta (5.0 nm) capping layer was grown after the substrate was cooled down to room temperature. The pressure of Ar working gas is 2.1 mTorr for all the layers. The structural features of $Pt_3Sn$ thin films were characterized by out-of-plane ($\theta$-$2\theta$ scan) x-ray diffraction (XRD) with Co-Kα radiation ($\lambda$ = 0.179 nm) using a Bruker D8 Discover system and by analytical electron microscopy using aberration-corrected FEI Titan G2 60-300 STEM equipped with super-X EDX detector. Cross-sectional samples for the STEM study were prepared by using a FEI focused-ion beam (FIB) system. The $Pt_3Sn$ and $Pt_3Sn_xFe_{1-x}$ samples were patterned into Hall bar devices by photolithography and Ar ion milling. And the electrical transport of $Pt_3Sn$ and $Pt_3Sn_xFe_{1-x}$ Hall bar devices was tested through DC setup measurement by utilizing a Physical Property Measurement System (Quantum Design, DynaCool).

### Device fabrication and electrical testing

The samples with the stack of MgO (001) sub./$Pt_3Sn$ (60.0 nm)/MgO (6.0 nm)/Ta (5.0 nm), MgO (001) sub./Pt (Mo) (2.0 nm)/$Pt_3Sn$ (60.0 nm)/MgO (6.0 nm)/Ta (5.0 nm) and MgO (001) sub./ $Pt_3Sn_xFe_{1-x}$ (60.0 nm)/MgO (6.0 nm)/Ta (5.0 nm) (numbers indicate the thickness in nm, same below), were prepared and patterned into Hall bar devices with 4~12-μm width and 144-μm length using an optical lithography process. The electrode with Ti (10 nm)/Au (150 nm) was deposited by CHA evaporator after etching the MgO (6.0 nm)/Ta (5.0 nm). Then temperature-dependent magnetoresistance, resistance, and Hall effect were tested by a physical property measurement system (PPMS) with a dc setup with Keithley's 2182 nanovoltmeter and 6221 current source.

The $Pt_3Sn$ (10.0 nm)/CoFeB (3.0 ~ 6.0 nm), Mo (2.0 nm)/$Pt_3Sn$ (10.0 nm)/CoFeB (3.0 ~ 6.0 nm), $Pt_3Sn_xFe_{1-x}$ (10.0 nm)/CoFeB (3.0 ~ 6.0 nm) and Pt (5.0 nm)/CoFeB (3.0 ~ 6.0 nm) samples were patterned into rectangular-shaped microstrips with dimensions of 5~20-μm width and 30-μm length by optical lithography and Ar ion milling. The electrode with Ti (10 nm)/Au (150 nm) was deposited by CHA evaporator. Symmetric coplanar waveguides in the ground-signal-ground (GSG) form were utilized for microwave injection into the $Pt_3Sn$ ($Pt_3Sn_xFe_{1-x}$)/CoFeB microstrips. A bias tee was used to inject microwave current and measure the resulting dc voltage at the same time. During the measurement, a microwave current with constant frequency (6 ~ 15 GHz) is injected while a magnetic field is swept at an angle of 45° with respect to the microstrips, and the output dc voltage is measured at each magnetic field with Keithley's 2182 nanovoltmeter.

**DFT calculations**

The total-energy electronic structure calculations were carried out using first-principles methods based on DFT. The generalized gradient approximation exchange-correlation potentials plus the projector augmented wave method for the electron-ion interaction was used (55), as implemented in Vienna *ab initio* simulation package code (56). All self-consistent calculations were performed with a plane-wave cutoff of 500 eV.

The geometric optimizations were carried out without any constraint until the force on each atom is less than 0.01 eV/A and the change of energy per cell is smaller than $10^{-5}$ eV. The Brillouin zone k-point sampling was set with a 21×21×21 Γ-centered Monkhorst-Pack grids. Wannier 90 package was used to fit the DFT band structures and calculate the SHC of $Pt_3Sn$ alloys (57). Fe doping is performed using the virtual crystal approximation method with a doping ratio of 5%, as suggested from experiments.

**Acknowledgments:** This project is supported by SMART, one of seven centers of nCORE, a Semiconductor Research Corporation program, sponsored by National Institute of Standards and Technology (NIST). T. P. and D.L.Z. were partly supported by ASCENT, one of six centers of JUMP, a Semiconductor Research Corporation program that is sponsored by MARCO and DARPA. Parts of this work were carried out in the Characterization Facility of the University of Minnesota, which receives partial support from the NSF through the MRSEC (Award Number DMR-2011401). Portions of this work were conducted in the Minnesota Nano Center, which is supported by the National Science Foundation Nano Coordinated Infrastructure Network (NNCI) under Award Number ECCS-2025124. J.P.W. and D.L.Z. are grateful for the useful discussions with former Nanomagnetism and Quantum Spintronics group members at UMN: Dr. Mahendra D.C. and Dr. Lakhan Bainsla.

**Author contributions:** J.P.W. initialized and planned the project. D.L.Z. and J.P.W. conceived the experiments. D.L.Z. designed and prepared all the samples. W.J. and T.L. performed the DFT calculations and proposed the model. H.Y. and K.A.M. carried out STEM and EDX studies. D.L.Z. T.P, Z.C. G.Y. and J.G.B. carried out the XRD measurements. D.L.Z. and O.J.B. patterned the Hall Bar devices for ST-FMR measurement. D.L.Z. carried out the ST-FMR measurements with Y.F.. P.S. contributed to the discussion of the crystalline phase and the band structure of $PtSn_4$ and $Pt_xSn_y$ materials in general. D.L.Z. and W.J. wrote the manuscript with J. P. W. and T. L. All the authors discussed the results and commented on the manuscript.

**Competing interests:** Authors declare no competing interests.

**Data and materials availability:** All data are available in the manuscript or the Supplementary Materials.

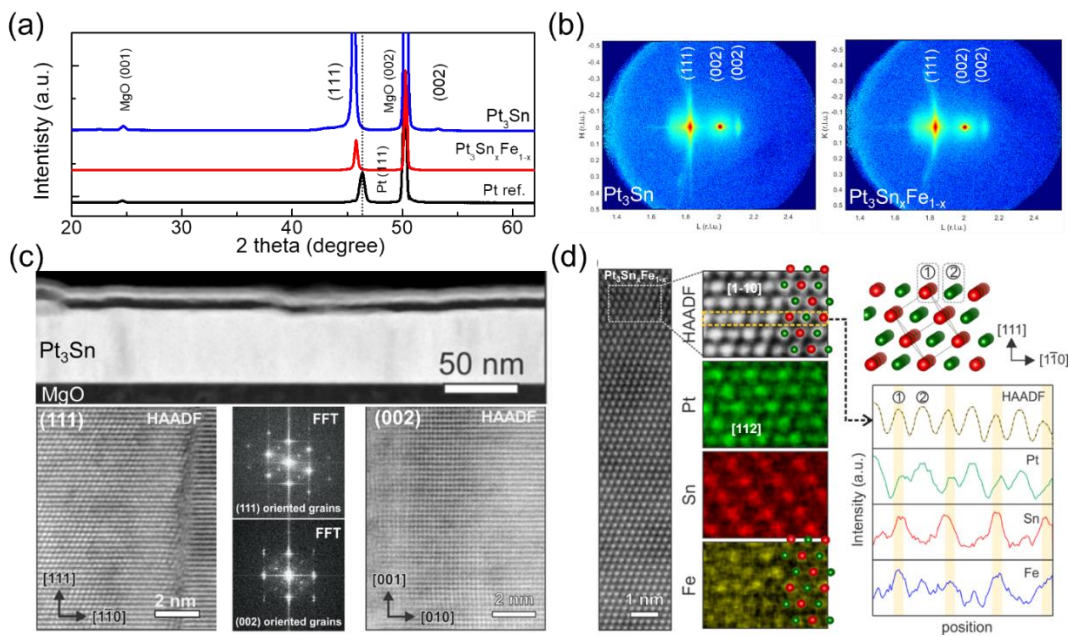

**Fig. 1. Crystalline structure of the Pt₃Sn samples.** (a) Specular ($\theta$-$2\theta$ scans) XRD patterns of the Pt reference, Pt$_3$Sn and Pt$_3$Sn$_x$Fe$_{1-x}$ thin films. (b) Reciprocal space maps (RSM) around the (002) Bragg reflection of the MgO substrate of the Pt$_3$Sn and Pt$_3$Sn$_x$Fe$_{1-x}$ thin films. Both the Pt reference and the Pt$_3$Sn and Pt$_3$Sn$_x$Fe$_{1-x}$ thin films grow epitaxially on the MgO substrate along the (111) direction. The XRD experiments show a small amount of (001) oriented grains. (c) HAADF-STEM images of the Pt$_3$Sn thin film on the MgO substrate. Low-magnification image (top panel) shows the Pt$_3$Sn film and capping layers with relatively uniform thicknesses. Atomic-resolution HAADF-STEM images obtained from (111) oriented (bottom-left) and (002) oriented (bottom-right) grains demonstrate their crystalline orientations. Fast Fourier transforms (FFTs) from the (111) and (002) oriented grains are also displayed (bottom-middle). (d) Atomic-resolution HAADF-STEM image and EDX elemental maps of the Pt$_3$Sn$_x$Fe$_{1-x}$. Schematic of the atomic structure is illustrated along with elemental line profiles, extracted from the region in the yellow-dashed line on the HAADF-STEM image.

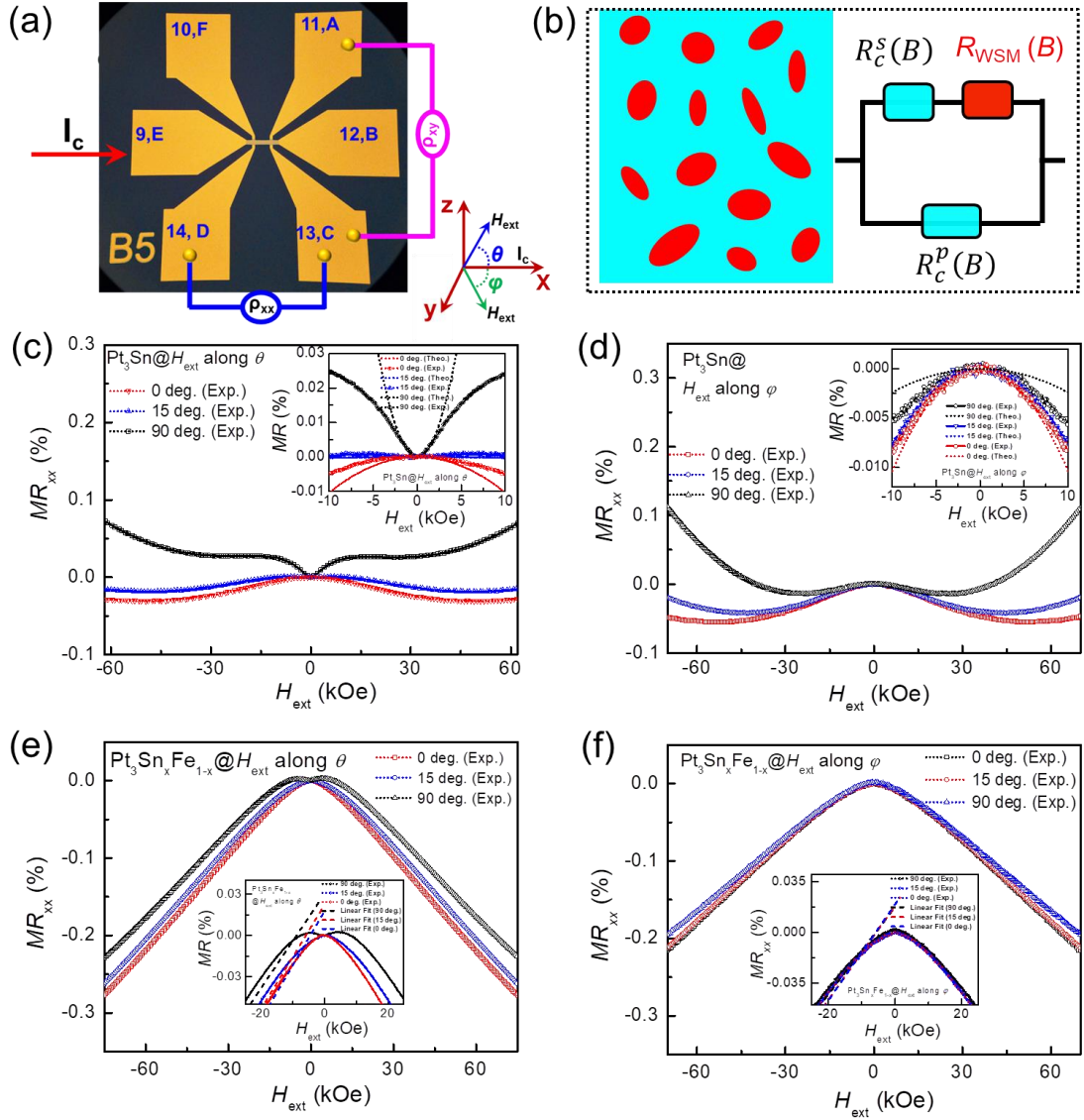

**Fig. 2. Characterization and physical origin of robust NLMR. (a)** The photo image of Hall bar devices used for electric-transport measurements. Where the current ($I_c$) is applied along x-axis, the external magnetic field ($H_{ext}$) rotates along $xz$ plane with angle $\theta$ (out-of-plane magnetoresistance ($MR_{xx}$)) and $xy$ plane with angle $\varphi$ (in-plane $MR_{xx}$), respectively. **(b)** Schematic of three-resistor model with conventional resistor (color blue) each connected in series ($R_c^s$) and in parallel ($R_c^p$) with the topological semimetal $R_{SM}$ (color red). The red areas represent contributions from topological semimetal (both DSM and WSM) and the blue areas denote normal metallic contribution. **(c)** and **(d)** The experimental measured and theoretically-fitted $MR_{xx}$ vs. $H_{ext}$ curves of $Pt_3Sn$ for $\theta$-angle dependence and $\varphi$-angle dependence, respectively. The theoretically-fitted $MR_{xx}$ vs. $H_{ext}$ curves based on three-resistor model are shown in the insert of **(c)** and **(d)**. **(e)** and **(f)** The experimental measured and theoretically-fitted $MR_{xx}$ vs. $H_{ext}$ curves of $Pt_3Sn_xFe_{1-x}$ for $\theta$-angle dependence and $\varphi$-angle dependence, respectively. The theoretically-fitted $MR_{xx}$ vs. $H_{ext}$ curves based on three-resistor model are shown in the insert of **(e)** and **(f)**.

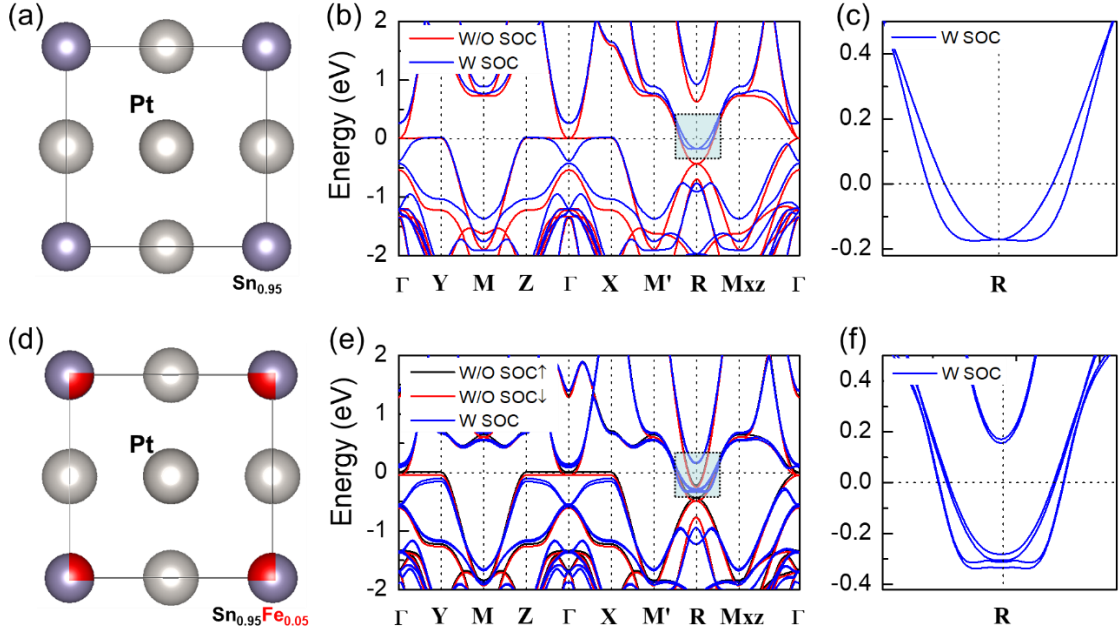

**Fig. 3. Theoretical calculation of topological properties.** (**a**) and (**b**) The crystal structure of pristine $Pt_3Sn$ and the band structure of pristine $Pt_3Sn$ that shows weak topological features. Blue and red lines represent band structures with and without spin orbit coupling, respectively. (**c**) Enlarged band structure with the Dirac nodes around the R point. (**d**) and (**e**) The crystal structure of pristine $Pt_3Sn_xFe_{1-x}$ and its band structure that shows clear spin-splitting and formation of TRS-broken WSM. Blue, black (spin up), and red (spin down) lines represent band structures with/without spin orbit coupling, respectively. (**f**) Enlarged band structure with the Weyl nodes around the R point that shows formation of TRS-broken WSM.

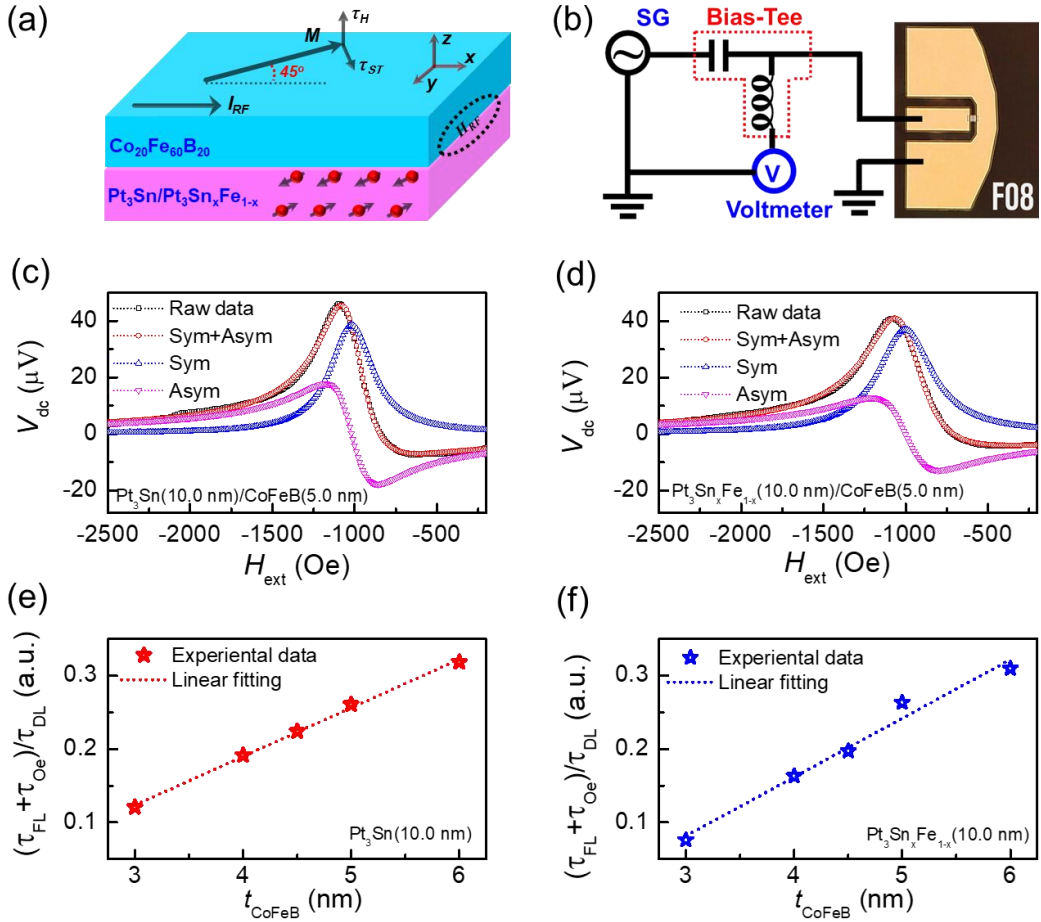

**Fig. 4. Spin torque efficiency.** **(a)** and **(b)** The schematic of the ST-FMR measurement and experimental setup with the microscopy of the device. **(c)** and **(d)** the room-temperature ST-FMR spectra for the Pt$_3$Sn (10.0 nm)/CoFeB (5.0 nm) and Pt$_3$Sn$_x$Fe$_{1-x}$ (10.0 nm)/CoFeB (5.0 nm) devices. **(e)** and **(f)** The ratio ($\tau_{FL} + \tau_{Oe}$)/$\tau_{AD}$ plotted against the thickness of the CoFeB layer. Through the linear fitting, spin torque efficiencies $\theta_{SH}$ of ~ 0.4 and ~ 0.38 are obtained for Pt$_3$Sn (10.0 nm) and Pt$_3$Sn$_x$Fe$_{1-x}$ (10.0 nm), respectively.